\documentclass[11pt]{article}
\def\fbin{\rm{fb^{-1}}}

\usepackage{graphicx}
\usepackage{epsfig}

\newcommand{\BABARPubYear}    {04}

\newcommand{\BABARConfNumber} {22}
\newcommand{\SLACPubNumber} {10632}

\input pubboard/babarsym

\long\def\inst#1{\par\nobreak\kern 4pt\nobreak
    {\it #1}\par\vskip 10pt plus 3pt minus 3pt}

\begin{document}
{\pagestyle{empty}

\begin{flushright}
%\babar\ Analysis Document \#982, Version 9 \\
\babar-CONF-\BABARPubYear/\BABARConfNumber \\
%\babar-PUB-\BABARPubYear/\BABARPubNumber \\
SLAC-PUB-\SLACPubNumber \\
%hep-ex/\LANLNumber \\
August 2004 \\
\end{flushright}

\par\vskip 5cm

% Title of the paper
\begin{center}
\Large \bf {Measurement of the Ratio of Branching Fractions of $\Xi_c^0$ Decays to $\Xi^-\pi^+$ and to $\Omega^-{\rm K}^+$} 
\end{center}
\bigskip

\begin{center}
\large The \babar\ Collaboration\\
\mbox{ }\\
\today
\end{center}
\bigskip \bigskip

% Abstract
%
% NB! If you update the text here, be sure to change the abstract.tex
% file too. That is a plain-latex version of the abstract for the eprint
% submission. No user-defined symbols allowed in it.
%
\begin{center}
\large \bf Abstract
\end{center}
We have analyzed 116 $\fbin$ of data collected by the \babar\ detector
for $\Xi_c^0$ production. In this paper we describe the
observation of $\Xi_c^0$ production from \ccbar continuum and from B decays,
with the $\Xi_c^0$  decaying into $\Xi^-\pi^+$ and $\Omega^- K^+$ modes.
The ratio of the branching fractions of the $\Xi_c^0$ decays into these two
final states measured in the \ccbar continuum is:
\begin{displaymath}
  \frac{ \mathcal{B}(\Xi_c^0 \rightarrow \Omega^- K^+) }{ \mathcal{B}(\Xi_c^0 \rightarrow \Xi^- \pi^+) } =  0.296 \pm 0.018~\mathrm{(stat.)} \pm 0.030~\mathrm{(sys.)} .
\end{displaymath}
All results in this note are preliminary.
\vfill
\begin{center}

Submitted to the 32$^{\rm nd}$ International Conference on High-Energy Physics, ICHEP 04,\\
16 August--22 August 2004, Beijing, China

\end{center}

\vspace{1.0cm}
\begin{center}
{\em Stanford Linear Accelerator Center, Stanford University, 
Stanford, CA 94309} \\ \vspace{0.1cm}\hrule\vspace{0.1cm}
Work supported in part by Department of Energy contract DE-AC03-76SF00515.
\end{center}

\newpage
} % end of pagestyle{empty}

% Input author list file
\begin{center}
\small

The \babar\ Collaboration,
\bigskip

%% author list as of 02-Jul-2004 (609 authors)
%
B.~Aubert,
R.~Barate,
D.~Boutigny,
F.~Couderc,
J.-M.~Gaillard,
A.~Hicheur,
Y.~Karyotakis,
J.~P.~Lees,
V.~Tisserand,
A.~Zghiche
\inst{Laboratoire de Physique des Particules, F-74941 Annecy-le-Vieux, France }
% ...

\end{center}\newpage

% The body of the paper starts here
\section{INTRODUCTION}
\label{sec:Introduction}

Little is known about charmed baryons today even though decades have passed 
since the discovery of charm. The high-luminosity $B$-factories 
present excellent opportunities to study the production and decay of
charmed baryons with high precision.

We present the observation of the $\Xi_c^0$ (csd)\footnote{
  All channels imply the charge conjugates as well, unless otherwise specified.
} charmed baryon in two decay modes:
\begin{eqnarray*}
     &&\Xi_c^0 \rightarrow \Omega^-\: K^+
  \\ &&\Xi_c^0 \rightarrow \Xi^- \: \pi^+  .
\end{eqnarray*}
The ratio of branching fractions of these decay modes has been
predicted to be approximately 0.32~\cite{ref:theory_prediction} 
using a spectator quark model calculation. This figure has a substantial
theoretical uncertainty. It has been measured previously by the CLEO
collaboration; their result was consistent with this prediction but
had a large statistical uncertainty~\cite{ref:cleo_paper}.

We measure the ratio of the branching fractions of 
$\Xi_c^0 \rightarrow \Omega^- K^+$ and $\Xi_c^0 \rightarrow \Xi^-\pi^+$
from the continuum production of $e^+ e^- \rightarrow \ccbar$,
where the hyperons are reconstructed through the following decays:
\begin{eqnarray*}
     &&\Xi^-   \rightarrow \Lambda \: \pi^-
  \\ &&\Omega^-   \rightarrow \Lambda \: K^- 
  \\ &&\Lambda \rightarrow p \: \pi^- .
\end{eqnarray*}
Since the two final states are topologically similar, quite a few systematic uncertainties
cancel in the ratio of the branching fractions. We also observe signals for
${ \Upsilon(4S) \rightarrow B\bar B \rightarrow \Xi_c^0 +  X}$
in both final states, where $X$ represents the rest of the event. 
Although copious production of $\Xi_c^0$ and $\Xi_c^+$ in $B$ decays has been
predicted~\cite{bib:theory_BtoXic_production}, this process has
been observed previously only by CLEO, with a significance of approximately
3~$\sigma$ in the $\Xi_c^0 \rightarrow \Xi^- \pi^+$ decay mode and
approximately 4~$\sigma$ in a related $\Xi_c^+$ decay mode~\cite{bib:cleo_BtoXic}.

% %\cite{Luo:2003pv}
% \bibitem{Luo:2003pv}
% Z.~Luo and J.~L.~Rosner,
% %``Final-state phases in B $\to$ baryon antibaryon decays,''
% Phys.\ Rev.\ D {\bf 67}, 094017 (2003)
% [arXiv:hep-ph/0302110].
% %%CITATION = HEP-PH 0302110;%%

\section{THE \babar\ DETECTOR AND DATASET}
\label{sec:babar}
The data for this analysis are collected with the \babar\ detector
at the \pep2\ asymmetric $e^+e^-$ collider; a total integrated luminosity of
116 $\fbin$ is used.
A five-layer silicon vertex detector (SVT) and a 40-layer drift chamber (DCH)
form the tracking system. The drift chamber is surrounded by the DIRC, a detector of
internally reflected Cherenkov light, which provides additional charged particle identification (PID). 
These are enclosed in a 
CsI(Tl) electromagnetic calorimeter (EMC). The detector assembly is embedded in a 1.5 T
superconducting magnet. Further details of the \babar\ detector are given elsewhere~\cite{ref:babar}.

The data collected are processed through the standard \babar\ reconstruction software. 
In the present analysis, we use
an integrated luminosity of $105.4 ~\fbin$ collected at the ${ \Upsilon(4S)}$ resonance,
and $10.7 ~\fbin$ collected below the ${ \Upsilon(4S)}$ threshold. We refer to these 
as the on-peak and the off-peak data samples, respectively.

\section{ANALYSIS METHOD}
\label{sec:Analysis}

The two decay modes $\Omega^- \rightarrow \Lambda K^-$ and 
$\Xi^- \rightarrow \Lambda \pi^-$ are topologically similar.
The hyperons involved are long-lived.
The $\Xi_c^0$ decays close to the production vertex: 
c$\tau  = 34^{+4}_{-2}~\mu$m~\cite{ref:PDGbook}.

\subsection{Selection of Events}
\label{sec:selection}

The $\Xi_c^0$ reconstruction takes place in three stages: 
\begin{itemize}
  \item{pre-selection of events containing a $\Lambda$,} 
   \item{pre-selection of events containing either $\Xi^-$ or $\Omega^-$ from the 
$\Lambda$ sample, and}
   \item{construction of $\Xi_c^0$ candidates.}
\end{itemize}

The $\Lambda$ is reconstructed by identifying a proton 
and combining it with an oppositely
charged track. $\Lambda$ candidates within a $3\sigma$ 
($\sigma$ is the fitted mass resolution) range
of the central value are then used for reconstruction of 
$\Xi^-$ and $\Omega^-$ by vertexing it with a negatively charged track; 
and the $\Lambda$ mass is 
constrained at the nominal value~\cite{ref:PDGbook}.
For $\Omega^-$ reconstruction, the $K^-$ is required to be identified as a kaon.

To improve signal-to-noise ratio, 
a minimum decay distance of 2.5 mm between the primary
vertex and the $\Xi^-$ decay vertex in the plane perpendicular to the beam direction is
required; for the $\Omega^-$, the required distance is 1.5 mm. In addition,
the ``signed'' flight distance\footnote{A ``signed'' flight 
length is where the displacement and the
momentum vector of the particle are required to be less than $90^\circ$ apart, i.e., 
${\bf\rm p\cdot r} >$ 0, where ${\bf\rm r}$ denotes the distance from the 
production point to the decay point of particle X in the xy-plane.}
between the $\Lambda$ and the $\Omega^-$ decay vertex is 
required to be at least 3~mm.

\begin{figure}
  \begin{center}
    \begin{tabular}{cc}
         \epsfig{file=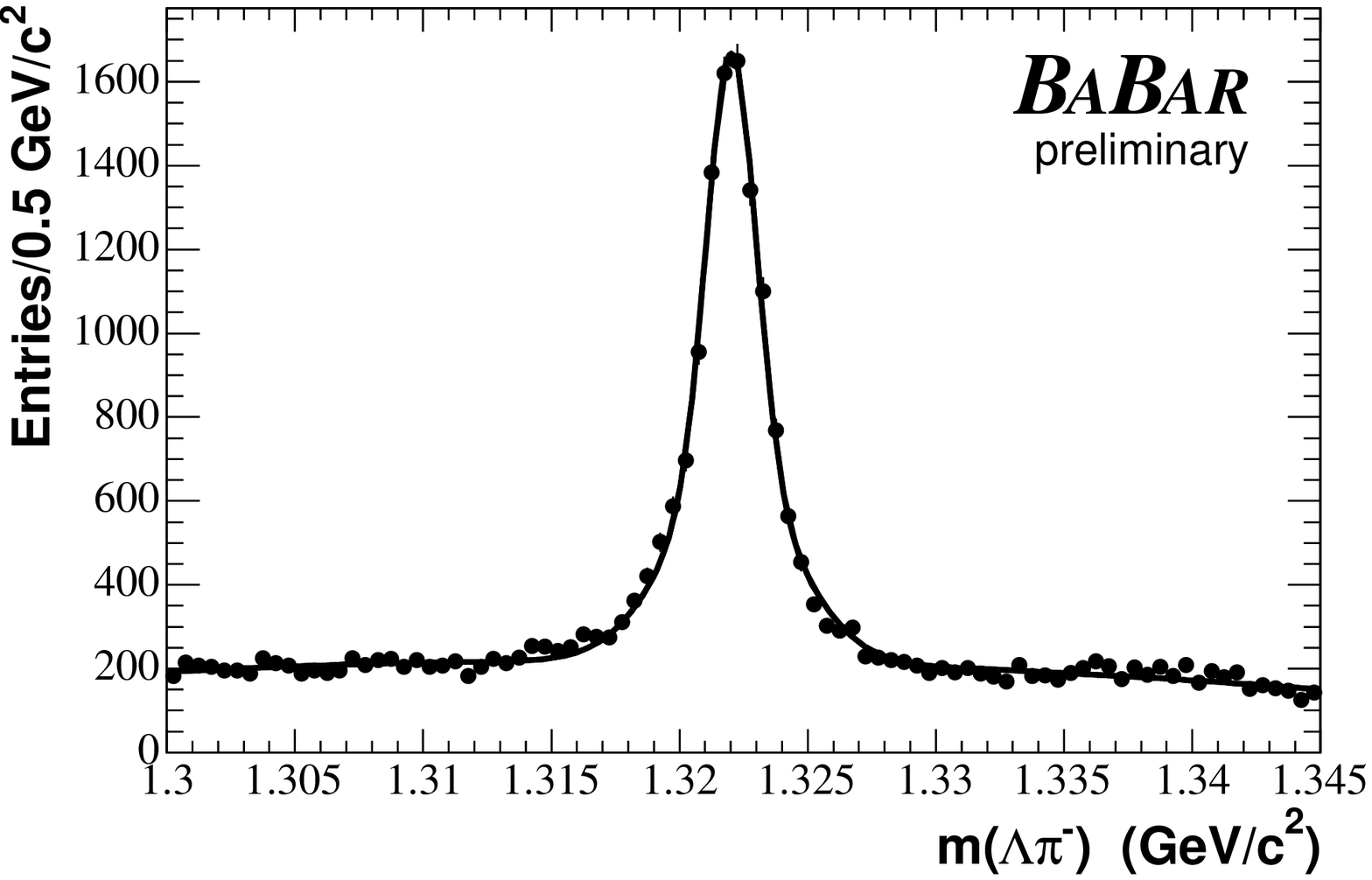, width=0.45\textwidth}
       & \epsfig{file=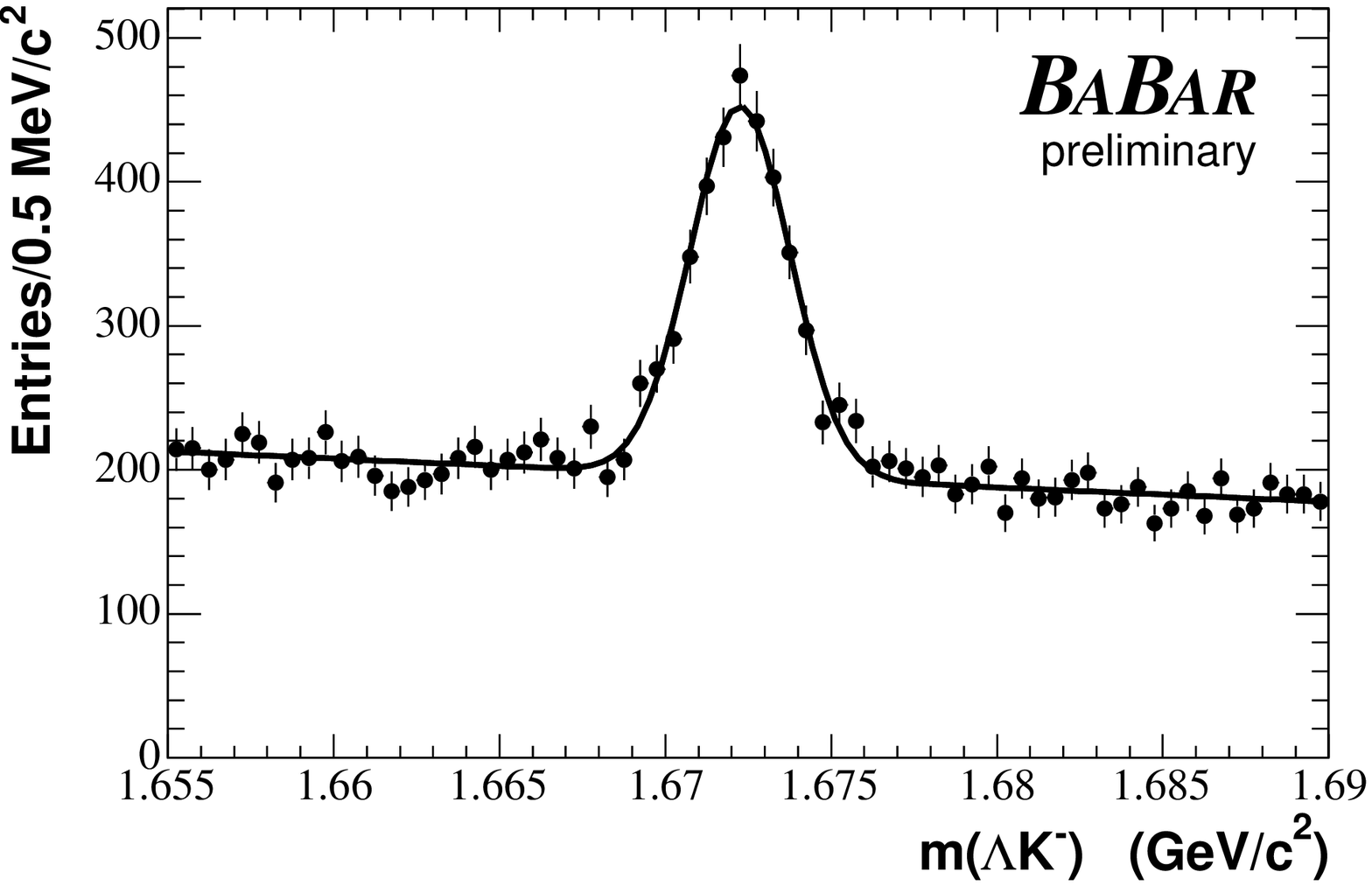, width=0.45\textwidth}
     \\ (a) & (b)
    \end{tabular}
  \end{center}
  \caption[Reconstructed and preselected $\Xi^-$ and $\Omega^-$ ]
          {Invariant masses of reconstructed and preselected (a) $\Xi^-$ and (b) $\Omega^-$ candidates 
           from subsamples of data.}
  \label{fig:XiOmega}
\end{figure}

Figures~\ref{fig:XiOmega} (a) and (b) show invariant mass distributions of
$\Lambda\pi^-$ and $\Lambda K^-$ respectively, from
subsamples of the data.
Superimposed on the plots are fits to a double Gaussian (single Gaussian)
for the $\Xi^-$($\Omega^-$) together with a linear background.
The fitted masses and resolutions of data and Monte Carlo are consistent
within known systematic effects. 

\subsection{$\Xi_c^0$ Reconstruction}
\label{sec:xic_reconstruction}

The selection criteria are finalized using subsamples
of the data as a precaution against selection bias.
The subsamples used are of size 20 $\fbin$ and 40 $\fbin$
for the $\Xi^-\pi^+$ and $\Omega^- K^+$ modes respectively.
The final results are obtained using the entire 116~\invfb
sample, including these subsamples.
In each case a 3$\sigma$ mass range around the 
central value is used.

\begin{figure}
  \begin{center}
    \begin{tabular}{cc}
      \epsfig{file=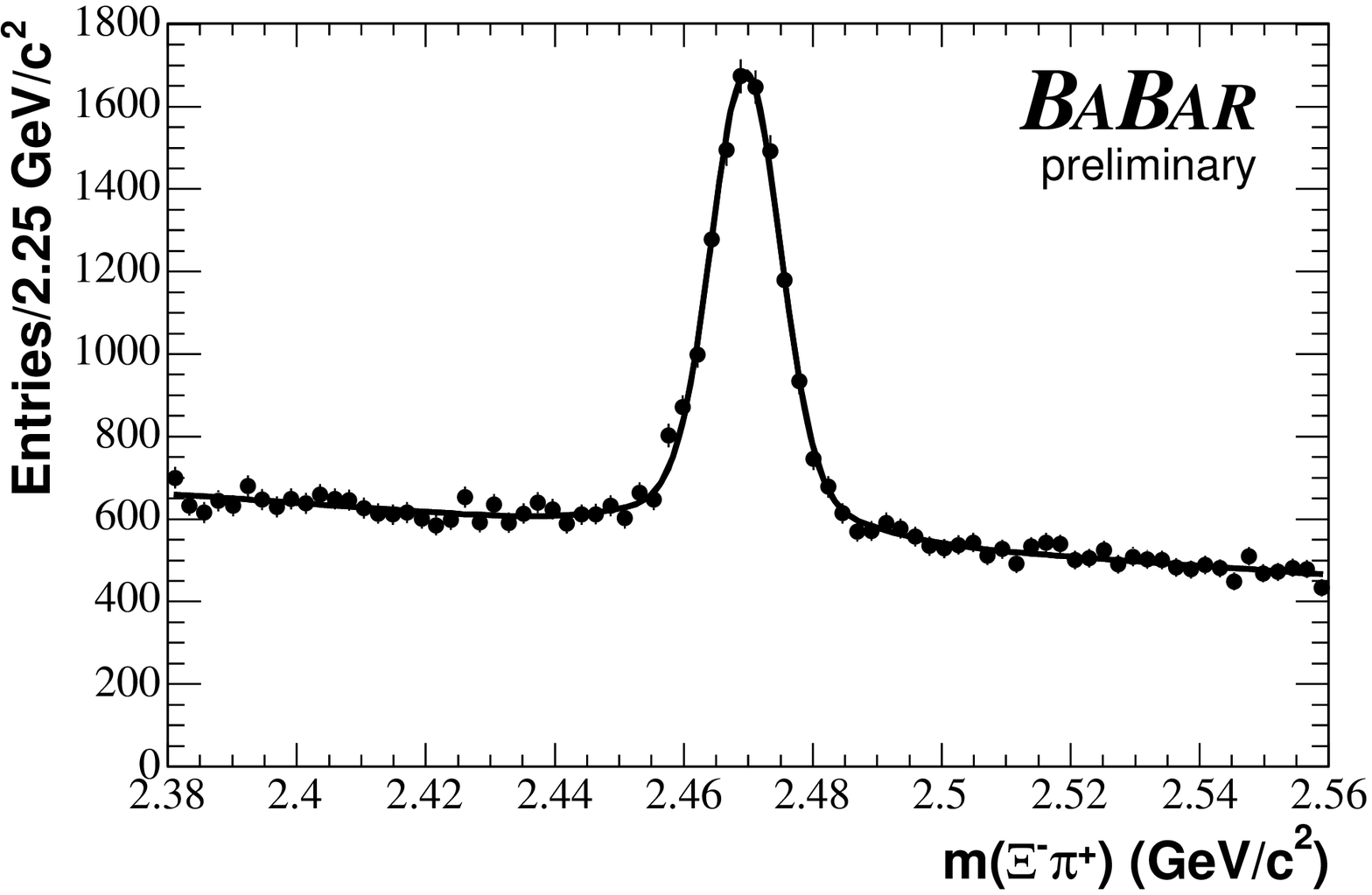, width=0.5\textwidth, angle=0}
      & \epsfig{file=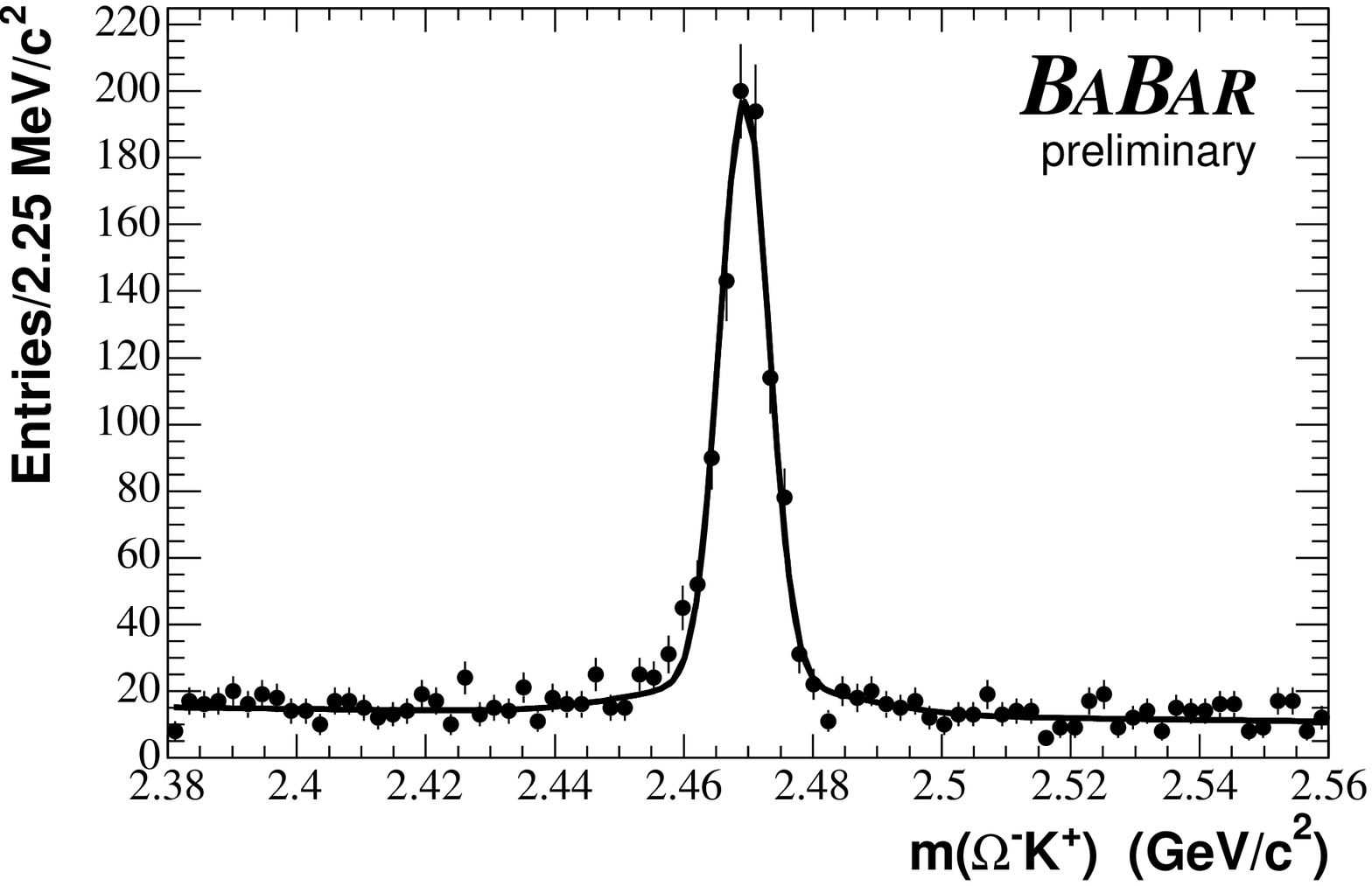, width=0.5\textwidth, angle=0}
      \\ (a) & (b)
      \\ \epsfig{file=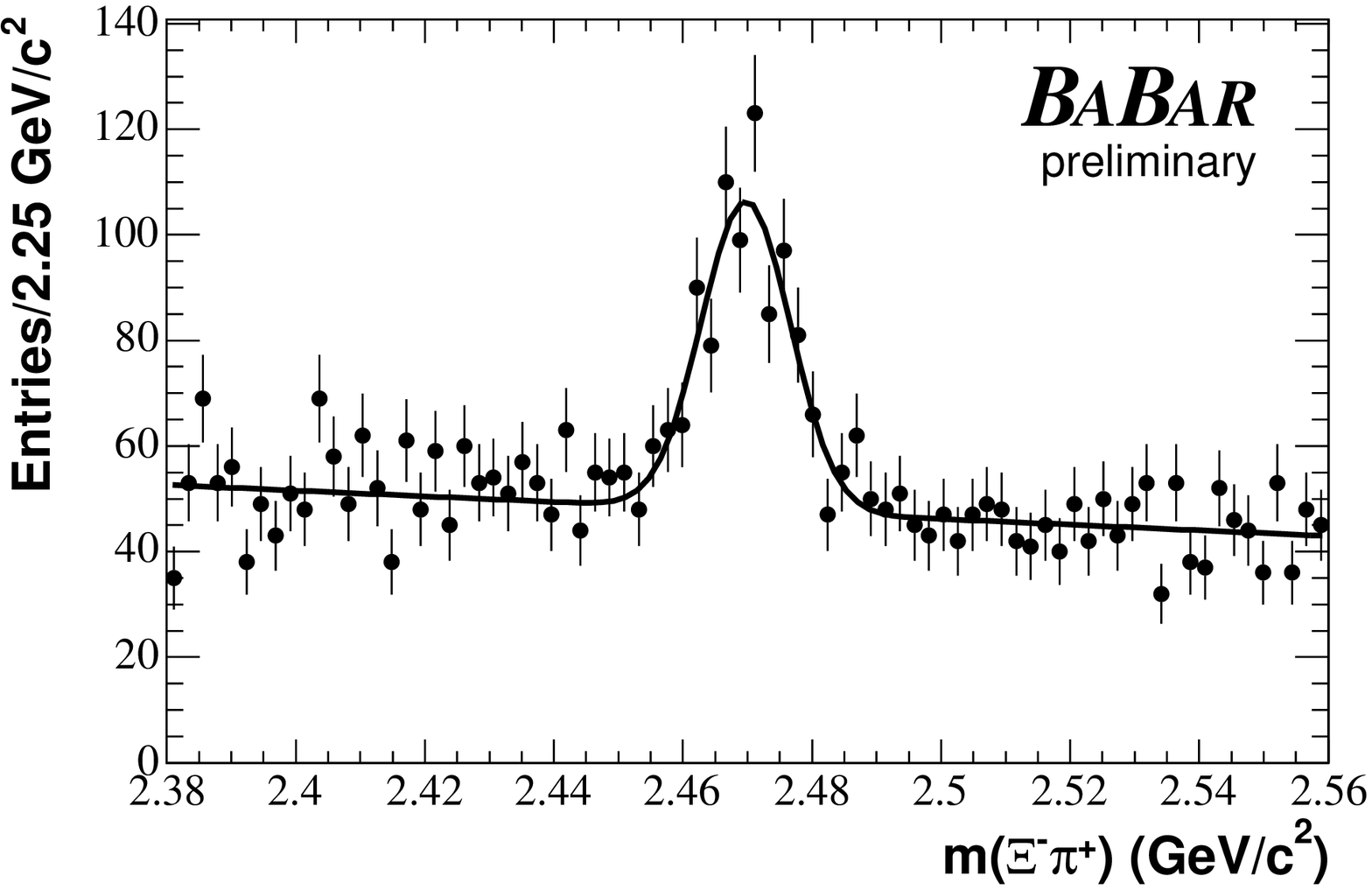, width=0.5\textwidth}
       & \epsfig{file=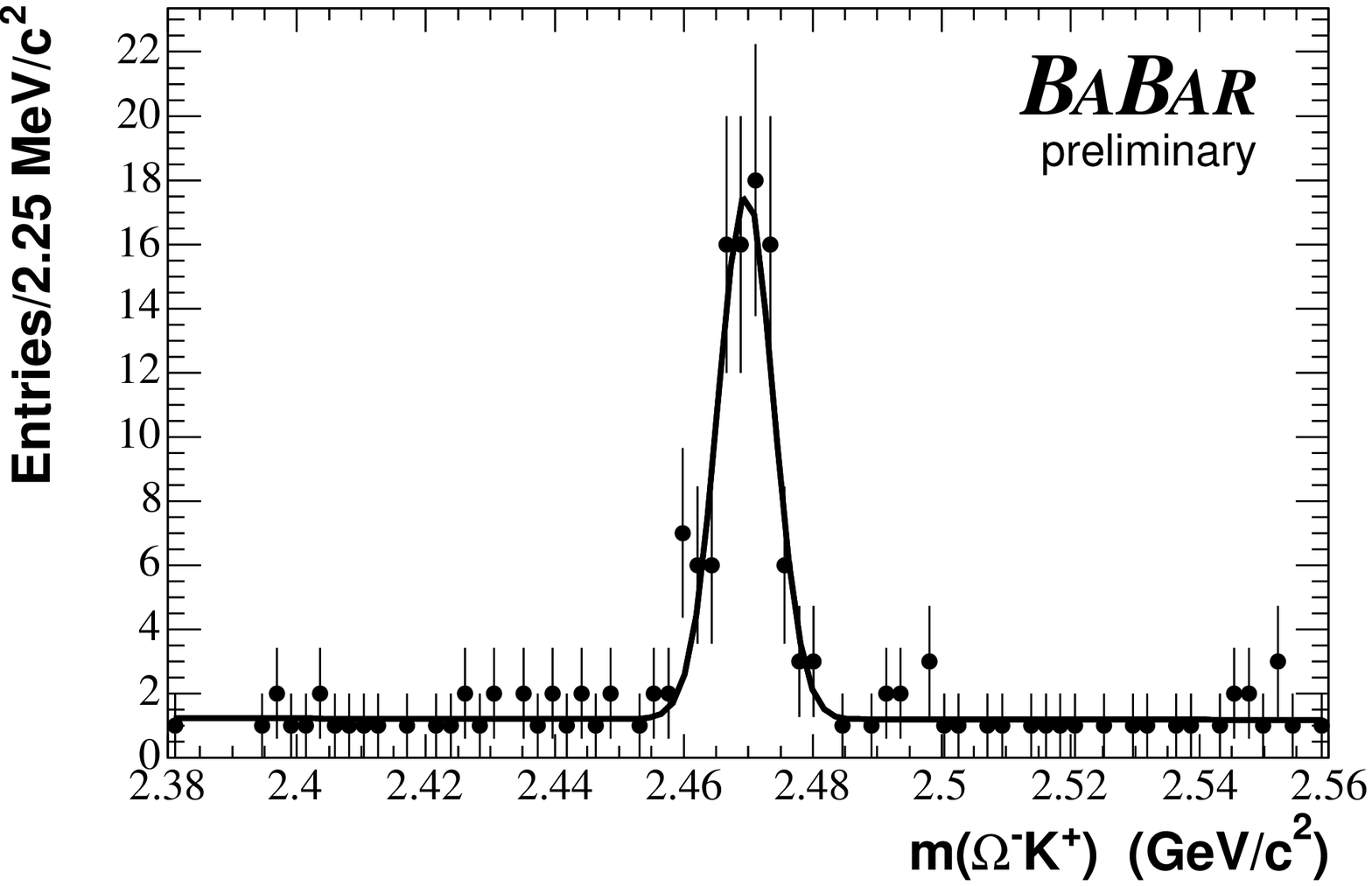, width=0.5\textwidth}
      \\ (c) & (d)
    \end{tabular}
  \end{center}
  \caption[Reconstructed $\Xi_c^0$ spectrum]
        {The invariant mass distributions for $\Xi_c^0$ candidates, shown for 
	  (a) $\Xi^- \pi^+$ in on-peak data,
	  (b) $\Omega^- K^+$ in on-peak data,
	  (c) $\Xi^- \pi^+$ in off-peak data, and
	  (d) $\Omega^- K^+$ in off-peak data.
        }
  \label{fig:Xicdata}
\end{figure}

Each resulting $\Xi^-$ candidate
is then vertexed with an oppositely charged pion 
for the $\Xi^-\pi^+$ final state. Likewise, each $\Omega^-$ candidate is vertexed with a
positively charged track identified as a kaon for the $\Omega^- K^+$ final state. 
The resulting invariant mass distributions for the $\Xi_c^0$ candidates from
the on-peak data sample are shown in Figures~\ref{fig:Xicdata}~(a) and~(b) for $\Xi^-\pi^+$ 
and $\Omega^- K^+$ combinations, respectively. 
The mass distributions from the off-peak data sample are shown in Figures~\ref{fig:Xicdata}~(c)
and~(d) again for $\Xi^-\pi^+$ and $\Omega^- K^+$ combinations, respectively.
A clear $\Xi_c^0$ peak is evident in all four spectra. 
The fitted distributions are superimposed on the plots.
In each case we use a single Gaussian shape on a linear background,
except for (b) in which a much better fit is obtained by using
a double Gaussian shape on a linear background.
The fit results are listed in Table~\ref{tab:xicparams}.

\begin{table}
  \caption{Fit results for $\Xi_c^0$.}
 \begin{center}\small
  \begin{tabular}{|l|c|c|}
\hline\hline
\multicolumn{1}{|c|}{Yield in} & $\Xi^- \pi^+$      &   $\Omega^- K^+$ \\
\hline\hline
On-peak Data     &  7614 $\pm$ 545 & 906 $\pm$ 54        \\
\hline
Off-peak Data    &  450 $\pm$ 39        &    78 $\pm$ 10         \\
\hline
On- and  off-peak & & \\
Data, $p^* >$ 1.8 GeV/c, & 4058 $\pm$ 319   &    655 $\pm$ 43                \\
in $\cos\theta^*$ range: & ($-0.8 \leq \cos\theta^* \leq 0.8$) & ($-0.8 \leq \cos\theta^* \leq 0.6$)\\
\hline
   \end{tabular}
  \end{center}
  \label{tab:xicparams}
\end{table}

\subsection{Simulation}

Events corresponding to the
  $ e^+e^- \rightarrow \ccbar \rightarrow \Xi_c^0 + X$
process are generated, with the $\Xi_c^0$ decays into the two desired decay modes.
PYTHIA~\cite{ref:pythia} is used
for the $\ccbar$ fragmentation and GEANT4~\cite{ref:geant4}
is used to simulate the detector response.
These events are
then reconstructed and the selection criteria applied.
Samples of 90,000 events for the $\Xi^-\pi^+$ final state and 60,000 events for the $\Omega^- K^+$
final state are generated. To investigate possible background contributions, 
generic $e^+ e^- \rightarrow \qqbar~\{u,d,s,c\}$
continuum Monte Carlo events are processed through the complete analysis 
program sequence. The $e^+ e^- \rightarrow \ccbar$ sample corresponds
to an integrated luminosity of 64~\invfb, and the combined
$\uubar, \ddbar, \ssbar$ sample corresponds to 33~\invfb.
In addition, 22,000 events are generated according to
${ \Upsilon(4S) \rightarrow B\bar B \rightarrow \Xi_c^0 + X} $, and processed through
the complete analysis chain.

\subsection{Background Contributions}  

We analyze the generic $\ccbar$ Monte Carlo events. No evidence of a peaking
background is observed. The distribution of the background events can be fitted with a
linear shape in the $\Xi^-\pi^+$ channel. For the $\Omega^- K^+$ channel,
the distribution of the reconstructed events is flat. The
events from generic $\uubar, \ddbar, \ssbar$ Monte Carlo do not show any peaking
either. 

\subsection{$\Xi_c^0$ Production from $\ccbar$ Continuum and from 
${ \Upsilon(4S) \rightarrow B\bar B \rightarrow \Xi_c^0 + X}$}

\begin{figure}
  \begin{center}
         \epsfig{file=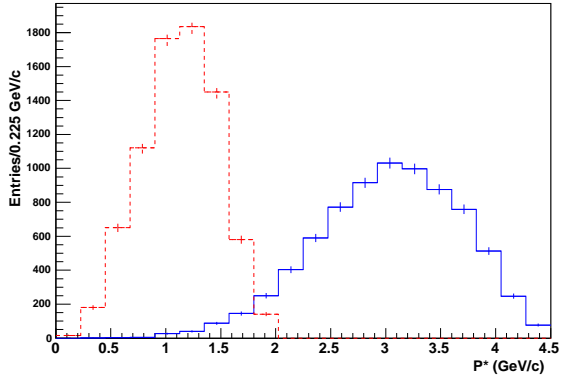, width=0.5\textwidth}
  \end{center}
  \caption[$p^*$ distribution from $B \rightarrow \Xi_c^0 + X$ 
    Monte Carlo.]
	  {$p^*$ distributions from reconstructed $\Xi_c^0$ in Monte Carlo.
	    The dashed red line shows $\Xi_c^0$ produced in B decays,
            and the solid blue line shows $\ccbar$ production of $\Xi_c^0$.
            Normalizations are arbitrary. No background is present.
          }
  \label{fig:p*xicMCrecon}
\end{figure}

The $B \rightarrow \Xi_c^0 + X$ Monte Carlo 
events are instructive in separating the $\Xi_c^0$ contribution
originating from B decays from those originating from 
the $\ccbar$ continuum production.
Figure~\ref{fig:p*xicMCrecon} shows 
the distribution of the momentum of the reconstructed $\Xi_c^0$'s
in the center-of-mass frame ($p^*$) from the Monte Carlo\footnote{
  The decay of the $B$ into $\Xi_c^0$ is modelled using 
  PYTHIA~\cite{ref:pythia} fragmentation.
}.
These are also ``truth-matched'', i.e., where the reconstructed
information matches the generated information.

The dashed red line shows the 
$p^*$ distribution
of $\Xi_c^0$'s originating from B decays and the
solid blue line shows that from $\ccbar$ continuum.
The normalizations in this figure are arbitrary.
The $p^*$ distribution from B decays does not 
extend beyond 2 GeV/c, purely from kinematics,
whereas the distribution from the
continuum peaks at much higher $p^*$ values.

\begin{figure}
  \begin{center}
    \begin{tabular}{cc}
      \\ \epsfig{file=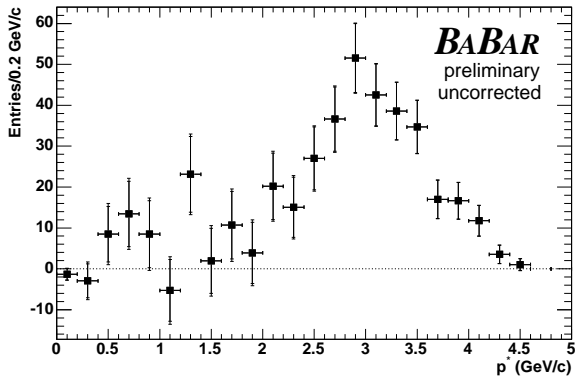, width=0.5\textwidth}
      &  \epsfig{file=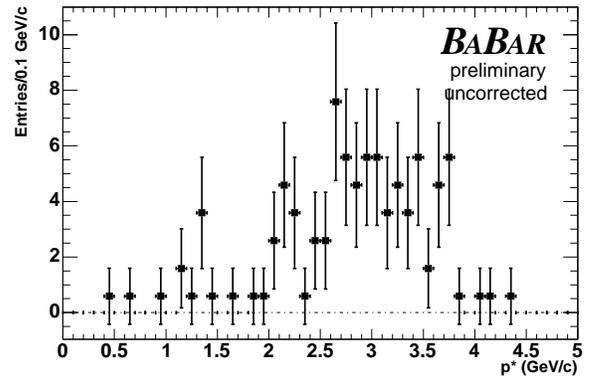, width=0.5\textwidth}
      \\ (a) & (b)
    \end{tabular}
  \end{center}
  \caption[$p^*$ distribution from the reconstructed $\Xi_c^0$ signal
          in (a) $\Xi^-\pi^+$ and (b) $\Omega^- K^+$ mode from off-peak
          data]
        {Sideband-subtracted $p^*$ distribution of reconstructed $\Xi_c^0$ candidates 
	  in off-peak data without efficiency correction in (a) $\Xi^-\pi^+$ and 
          (b) $\Omega^- K^+$ mode.
	  Most of the signal is produced at higher $p^*$ as expected.
	}
\label{fig:Xicoffpeak-p*}
\end{figure}

The $p^*$ distributions from the $\Xi_c^0$ signal regions for the off-peak data 
are shown without any efficiency correction in 
Figures~\ref{fig:Xicoffpeak-p*} (a) and (b),
for the $\Xi^-\pi^+$ and the
$\Omega^- K^+$ final states, respectively.
These data are collected below $\bbbar$ production
threshold, and therefore represent $\Xi_c^0$ production from continuum only. 
The background under the $\Xi_c^0$ signal in the data is estimated from the sidebands
in the reconstructed $\Xi_c^0$ mass spectrum and then removed. 
These $p^*$ distributions, peaked around 3 GeV/c, clearly indicate $\Xi_c^0$ 
production from $\ccbar$ continuum. 

\begin{figure}
  \begin{center}
    \begin{tabular}{cc}
       \epsfig{file=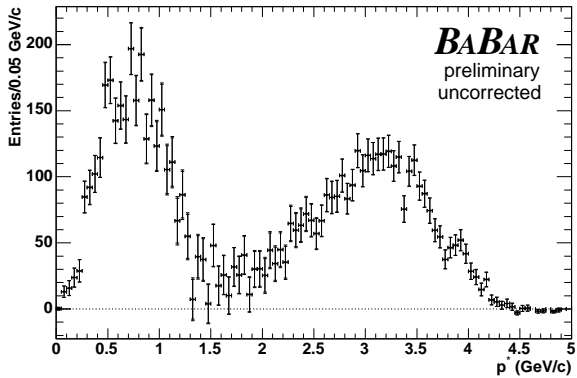, width=0.5\textwidth, angle=0} 
    &  \epsfig{file=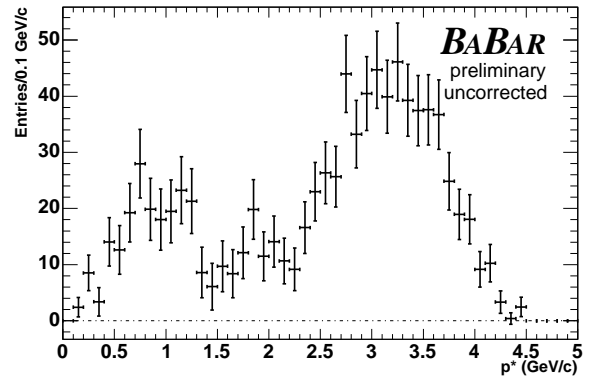, width=0.5\textwidth, angle=0} 
      \\ (a) & (b)
    \end{tabular}
  \end{center}
  \caption[$p^*$ distribution from $\Xi_c^0$ signal 
          from on-peak data in (a) $\Xi^-\pi^+$ and (b)  $\Omega^- K^+$ mode]
          {Sideband-subtracted $p^*$ distribution from $\Xi_c^0$ candidates in on-peak data without efficiency
           correction in (a) $\Xi^-\pi^+$ and (b)  $\Omega^- K^+$ mode.
            The lower peak below $p^* < 1.5$ GeV/c is primarily from
            the $\Xi_c^0$ production from B decays
            as evident from Figure~\ref{fig:p*xicMCrecon} and Figure~\ref{fig:Xicoffpeak-p*}. 
        }
  \label{fig:Xicdata_onpeakp*}
\end{figure}

Figures~\ref{fig:Xicdata_onpeakp*} (a) and (b) show the $p^*$ distribution in on-peak data
from the $\Xi_c^0$ candidates in the $\Xi^- \pi^+$ and $\Omega^- K^+$ final states, respectively, 
after background
subtraction, again without any efficiency correction. The peaks below 1.5 GeV/c in both
plots clearly represent $\Xi_c^0$ production from B decays, as evident
from Figures~\ref{fig:p*xicMCrecon} and~\ref{fig:Xicoffpeak-p*}.

\subsection{Analysis and Efficiency Correction for $\ccbar \rightarrow \Xi_c^0 + X$}

For further analysis the on- and off-peak data samples are combined;
to isolate the $\ccbar$ production of $\Xi_c^0$, events with
$p^* > 1.8$ GeV/c are selected. In order to avoid large fluctuations
from the edges of the phase-space and detector acceptance effects,
we also require $ -0.8 \leq \cos \theta^* \leq 0.8$ for the $\Xi^-\pi^+$ mode
and $ -0.8 \leq \cos \theta^* \leq 0.6$ the for $\Omega^- K^+$ mode, where
$\theta^*$ is the polar angle of the $\Xi_c^0$ candidate with respect to the collision axis
in the center-of-mass frame. 

\begin{figure}
  \begin{center}
    \begin{tabular}{cc}
         \epsfig{file=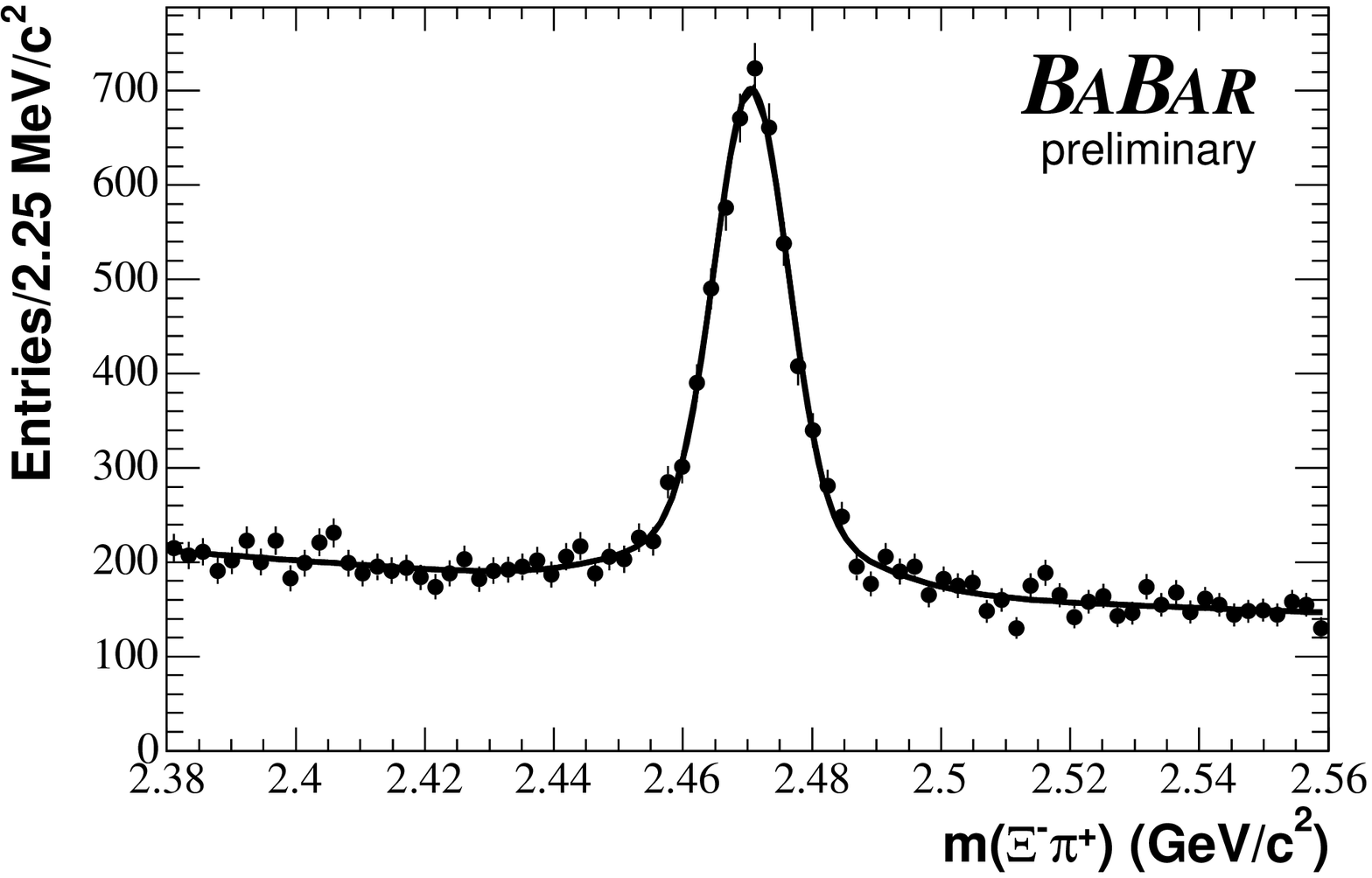, width=0.5\textwidth}
       & \epsfig{file=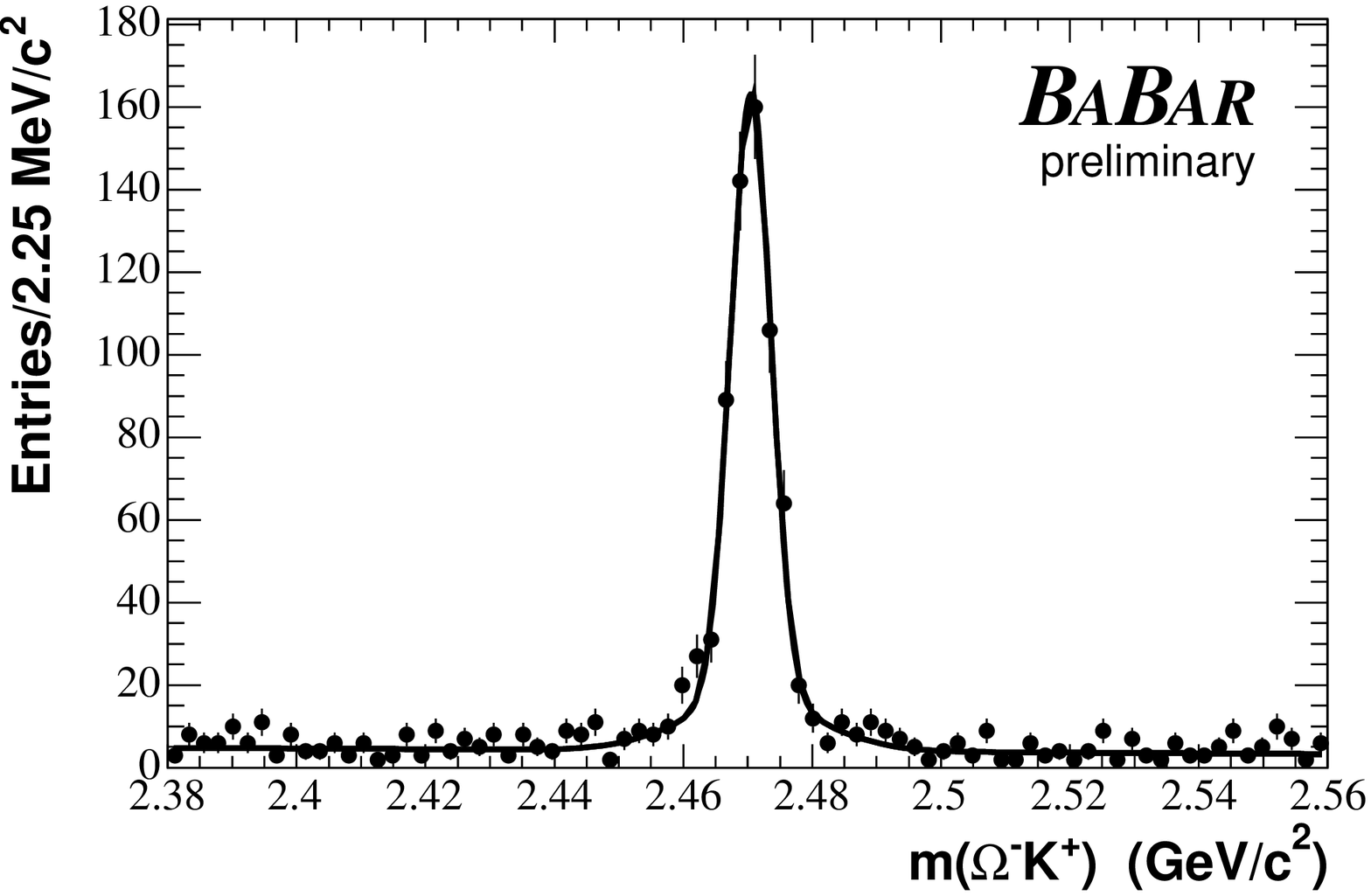, width=0.5\textwidth}
     \\ (a) & (b)
    \end{tabular}
  \end{center}
  \caption[$\Xi_c^0$ mass spectra from $\Xi^-\pi^+$ and $\Omega^- K^+$ with $p^* >$ 1.8 GeV/c and cos$\theta^*$ cut.]
          {Invariant mass spectra for $\Xi_c^0$ with $p^* >$ 1.8 GeV/c from (a) $\Xi^-\pi^+$ final state with 
           $ -0.8 \leq \cos \theta^* \leq 0.8$, and (b) $\Omega^- K^+$ final state with 
           $ -0.8 \leq \cos \theta^* \leq 0.6$
           using on- and off-peak data. Both spectra are fitted with a double Gaussian for signal and a linear background shape.
          }
  \label{fig:Xicmass_pcut_coscut}
\end{figure}

Figures~\ref{fig:Xicmass_pcut_coscut} (a) and (b) show the  
$\Xi_c^0$ invariant mass spectra in these $\cos \theta^*$ ranges 
with $p^* > 1.8 $ GeV/c for the combined on- and off-peak data sample, 
for $\Xi^-\pi^+$ and $\Omega^- K^+$ decay modes, respectively.
The fit results are presented in Table~\ref{tab:xicparams}. 

The efficiency is calculated from signal Monte Carlo events as a
function of  $p^*$ and cos$\theta^*$ in each of the two decay modes.
For each decay mode, a fifteen-parameter two-dimensional fit gives a smooth parameterization
of the efficiency with small statistical unceratinty. We then correct the data distribution by
weighting each event in the spectrum inversely by its efficiency according
to the event's position in ($p^*$, cos$\theta^*$) space.
After correcting for efficiency we obtain $ 19375 \pm 393$ events
in the $\Xi^-\pi^+$ mode and $ 4866 \pm 283$ events in the $\Omega^- K^+$ mode\footnote{
The $\Lambda$ branching fraction is not taken into account in these numbers, since this cancels
in the ratio.}.

The angular distribution of the data is well-described by a
$(1 + \cos^2\theta^*)$ function; we use this to estimate the
fractions of the $\Xi_c^0$ which are expected to lie in the selected
angular regions for each mode. Extrapolating from the fiducial
region into the full range $-1 \leq \cos\theta^* \leq +1$,
we obtain the total numbers of signal events for $p > 1.8$~\gevc are
$ 26621 \pm 540 $ and  $7874 \pm 458$
for the $\Xi^- \pi^+$ and $\Omega^- K^+$ modes, respectively.
We thus obtain the ratio of branching fractions:
   
\begin{displaymath}
  \frac{ B(\Xi_c^0 \rightarrow \Omega^- K^+)}{B(\Xi_c^0 \rightarrow \Xi^- \pi^+) } =  0.296 \pm 0.018~\mathrm{(stat.)} .
\end{displaymath}

\section{STUDY OF SYSTEMATIC UNCERTAINTIES}
\label{sec:Systematics}

We evaluate several sources of systematic uncertainties,
described below and summarized in Table~\ref{tab:systematics}.
Adding all of these uncertainties\footnote{
  No baryon polarization is considered in the present analysis and any
  systematic uncertainty due to this is neglected.
} in quadrature, we obtain a total absolute systematic uncertainty of 0.030
on the ratio of the branching fractions.  

\begin{itemize}
  \item{We vary the signal shape, the
    background shape, and the fit range,
    and use a simple event-counting method.
    From the deviations observed, we assign
    systematic uncertainties for the use of a binned fit
    and for the particular technique used, adding them
    in quadrature.
  }
  \item{We repeat the analysis with
    (a) a parameterization of the efficiency with a similar function
    with nine parameters, and (b) a simple efficiency calculation in
    two dimensional bins from Monte Carlo. The discrepancy observed
    between the main result and the result from (a) is
    assigned as the systematic uncertainty.
  }
  \item{We vary the range in $\cos\theta^*$ used for the two final states.
    We also vary the $\cos\theta^*$ distribution used for the extrapolation.
    The combined systematic uncertainty is taken to be the
    sum in quadrature of the uncertainty due to using different
    $\cos\theta^*$ ranges for the two modes and the uncertainty due to the 
    choice of extrapolation function.
  }
  \item{We take into account an uncertainty due to the finite size of the
    Monte Carlo sample used to estimate the efficiencies.}
  \item{Approximately 1\% of selected events contain multiple candidates
    in the $\Xi_c^0$ signal range with one or
    more tracks in common. We retain all such candidates, and
    therefore assign a systematic uncertainty in case these form
    a peaking background.}
  \item{$\Xi_c^0$ and $\bar{\Xi_c^0}$ are studied separately;
    the ratios are found to be consistent. We assign the difference
    as the systematic uncertainty due to detector charge asymmetry.}
  \item{We assign an uncertainty of 1\% on the efficiency for each track required to be identified as a kaon.}
  \item{The uncertainty in the branching fraction of $\Omega^-$, $(67.8 \pm 0.7)$\%~\cite{ref:PDGbook},
    is included.}
\end{itemize}

We make the following additional checks:
\begin{itemize}
\item{The desired ratio of the branching fractions is calculated from off-peak data only, and is
  measured to be $0.259 \pm 0.044$, consistent with the main result.}
\item{The data are divided up into three $p^*$ ranges: (1.8--2.7) GeV/c, (2.7--3.6) GeV/c, 
  and (3.6--4.5) GeV/c; the yields and ratios calculated for each range are $0.269 \pm 0.030$,
  $0.295 \pm 0.019$, and $0.263 \pm 0.031$, respectively. These are consistent 
  with being independent of $p^*$ within statistical uncertainties.}
\end{itemize}

\begin{table}
 \caption{Systematic uncertainties.}
 \begin{center}
   \begin{tabular}{|l|c|}
  \hline\hline
  \multicolumn{1}{|c|}{Source} & Uncertainty \\ \hline\hline
  Fits to mass spectrum & 0.019 \\
  Efficiency            & 0.015 \\
  $\cos\theta^*$ Distribution & 0.016 \\
  Limited Monte Carlo Statistics &  0.004 \\
  Multiple candidates   & 0.004 \\
  Charge asymmetry      & 0.001 \\ 
  Particle ID           & 0.006 \\ 
  $\Omega^-$ branching fraction & 0.003 \\ 
\hline
  Total systematic uncertainty     & 0.030 \\ \hline
   \end{tabular}
   \end{center}
 \label{tab:systematics}
\end{table}

\section{PHYSICS RESULTS AND SUMMARY}
\label{sec:Physics}

In summary, we observe $\Xi_c^0$ production from the 
\ccbar continuum and from ${\Upsilon(4S) \rightarrow B\bar B \rightarrow \Xi_c^0 + X}$ 
using the \babar\ detector at SLAC. 
This represents the first observation of $\Xi_c^0 \rightarrow \Omega^- K^+$ in $B$ decays.
We present a preliminary measurement of the ratio of 
branching fractions of $\Xi_c^0$ to $\Omega^- K^+$ 
and $\Xi^-\pi^+$, determined using the \ccbar continuum data:
\begin{displaymath}
  \frac{B(\Xi_c^0 \rightarrow \Omega^- K^+)}{B(\Xi_c^0 \rightarrow \Xi^- \pi^+)} =  0.296 \pm 0.018~\mathrm{(stat.)} \pm 0.030~\mathrm{(sys.)} .
\end{displaymath}
This represents a significant improvement on the
existing value of $(0.50 \pm 0.21 \pm 0.05)$~\cite{ref:cleo_paper}.

\section{ACKNOWLEDGMENTS}
\label{sec:Acknowledgments}

We are grateful for the 
extraordinary contributions of our \pep2\ colleagues in
achieving the excellent luminosity and machine conditions
that have made this work possible.
The success of this project also relies critically on the 
expertise and dedication of the computing organizations that 
support \babar.
The collaborating institutions wish to thank 
SLAC for its support and the kind hospitality extended to them. 
This work is supported by the
US Department of Energy
and National Science Foundation, the
Natural Sciences and Engineering Research Council (Canada),
Institute of High Energy Physics (China), the
Commissariat \`a l'Energie Atomique and
Institut National de Physique Nucl\'eaire et de Physique des Particules
(France), the
Bundesministerium f\"ur Bildung und Forschung and
Deutsche Forschungsgemeinschaft
(Germany), the
Istituto Nazionale di Fisica Nucleare (Italy),
the Foundation for Fundamental Research on Matter (The Netherlands),
the Research Council of Norway, the
Ministry of Science and Technology of the Russian Federation, and the
Particle Physics and Astronomy Research Council (United Kingdom). 
Individuals have received support from 
CONACyT (Mexico),
the A. P. Sloan Foundation, 
the Research Corporation,
and the Alexander von Humboldt Foundation.

\end{document}